\documentclass[prl,twocolumn,amsmath,amssymb,superscriptaddress,noshowpacs]{revtex4}

\usepackage{graphicx}
\usepackage{dcolumn}
\usepackage{bm}
\usepackage{color}

\usepackage{amsmath}
\usepackage{amssymb}

\begin{document}

\title{Charge order, dynamics, and magneto-structural transition in multiferroic LuFe$_2$O$_4$}

\author{X.S. Xu}
\affiliation{Department of Chemistry, University of Tennessee,
Knoxville, TN 37996, USA}
\author{{M. Angst}}
\affiliation{Oak Ridge National Laboratory,  Oak Ridge, TN 37831,
USA}
\affiliation{Institut f\"ur Festk\"orperforschung,
Forschungszentrum J\"ulich GmbH, D-52425 J\"ulich, Germany}
\author{T.V. Brinzari}
\affiliation{Department of Chemistry, University of Tennessee,
Knoxville, TN 37996, USA}
\author{R.P. Hermann}
\affiliation{Institut f\"ur Festk\"orperforschung, Forschungszentrum
J\"ulich GmbH, D-52425 J\"ulich, Germany} \affiliation{Department of
Physics, B5, Universit\'e de Li\`ege, B-4000 Sart-Tilman, Belgium}
\author{J.L. Musfeldt}
\affiliation{Department of Chemistry, University of Tennessee,
Knoxville, TN 37996, USA}
\author{A.D. Christianson}
\affiliation{Oak Ridge National Laboratory,  Oak Ridge, TN 37831,
USA}
\author{D. Mandrus}
\affiliation{Oak Ridge National Laboratory,  Oak Ridge, TN 37831,
USA} \affiliation{Department of Physics, University of Tennessee,
Knoxville, TN 37996}
\author{B.C. Sales}
\affiliation{Oak Ridge National Laboratory,  Oak Ridge, TN 37831,
USA}
\author{S. McGill}
\affiliation{National High Magnetic Field Laboratory, Tallahassee, FL, 32310, USA}
\author{J.-W. Kim}
\affiliation{Ames Laboratory, Ames, IA 50010, USA}
\author{Z. Islam}
\affiliation{Advanced Photon Source, Argonne National Laboratory,
Argonne, IL 60439, USA}

\date{\today}

\begin{abstract}

We investigated the series of temperature and field-driven
transitions in LuFe$_2$O$_4$ by optical and M\"{o}ssbauer
spectroscopies, magnetization, and x-ray scattering in order to
understand the interplay between charge, structure, and magnetism in
this multiferroic material. We demonstrate that charge fluctuation
has  an onset well below  the charge ordering transition, supporting
the ``order by fluctuation'' mechanism for the development of charge
order superstructure. Bragg splitting and large magneto optical contrast suggest a low
temperature monoclinic distortion that can be driven by both
temperature and magnetic field.

\end{abstract}



\maketitle Complex oxides take advantage of the unique and flexible
properties of transition metal centers to govern bonding and local
structure. Further, the delicate interplay between  charge,
structure, and magnetism yields important consequences for
functionality and cross-coupling. Iron-based materials such as
multiferroic BiFeO$_3$ and LuFe$_2$O$_4$ \cite{Kadomtseva2004,
Ikeda2005}, bistable photomagnetic systems such as Prussian blues
and related derivatives \cite{Shimamoto2002Berlinguette2004}, dilute
magnetic semiconductors \cite{Manyala2008}, and the new family
LaFeAsO$_{1-x}$F$_x$ of superconductors \cite{Kamihara2008} have
attracted recent attention. In this Letter, we focus
on LuFe$_2$O$_4$, a frustrated system with a series of 
phase transitions that give rise to
electronically-driven multiferroicity \cite{vandenBrink2008}.

LuFe$_2$O$_4$ has a layered structure with Fe-containing double
layers of triangular connectivity. Three-dimensional
Fe$^{2+}$/Fe$^{3+}$ charge order (CO) occurs below 320 K ($T_{CO}$).
This is followed by ferrimagnetic order below 240 K ($T_N$)
\cite{Ikeda2005,Xiang2007,Angst08}.  An additional low temperature
magnetic phase transition has recently been reported at 175 K
($T_{LT}$) \cite{Christianson08}.
The CO has a
 so-called
$\sqrt{3}$$\times$$\sqrt{3}$ superstructure
\cite{Xiang2007,Angst08}.  Because of the mixed valent iron centers
and frustrated triangular
lattice, the Fe$^{2+}$ and Fe$^{3+}$ populations are different within the 
double layer, an effect that renders an intrinsic polarization
\cite{Ikeda2005,Angst08}.
The charge ordering mechanism is thus central to understanding the
unusual physical properties of this 
multiferroic.

To elucidate the charge excitations and understand how they
correlate with structure and magnetism, we measured the optical and
M\"{o}ssbauer spectra, magnetization, and x-ray scattering of LuFe$_2$O$_4$. 
We compare our comprehensive results to recent electronic structure
calculations \cite{Xiang2007} and to spectral data on classical
magnetite \cite{Gasparov2000}. We demonstrate that strong
Fe$^{2+}\!\rightarrow$Fe$^{3+}$ charge fluctuation persists even in
charge ordered states characterized by superstructure reflections,
and it persists down to $T_{LT}$ below which Bragg splitting
indicates that strong monoclinic distortions occur. These
observations are consistent with the ``order by fluctuation''
mechanism \cite{Nagano2007}, in which case $\sqrt{3}$$\times$$\sqrt{3}$
CO is preferable for  entropy reasons 
and stabilized by the charge fluctuation in this geometrically
frustrated system. As in magnetite, we analyze the results in terms
of a polaron picture, extracting a large effective mass for the
charge carriers. On the other hand, Fe$^{2+}$ on-site excitations
are sensitive to the magnetic transition at $T_{LT}$ and display a
sizable magneto-optical effect. Combining our spectral, magnetic,
and structural data, we generate an $H$-$T$ phase diagram and show
that the transition at $T_{LT}$ can also be driven by a magnetic
field.
These results demonstrate the important interplay between charge,
structure, and magnetism.

All experiments were conducted on floating-zone-grown LuFe$_2$O$_4$
single crystals from the same batch as those used in Refs.
\cite{Christianson08,Angst08}. Near-normal reflectance measurements
were carried out on $ab$ plane samples
 employing a series of spectrometers covering a wide range of energy
 (30 meV - 6.5 eV),
  temperature (4  - 540 K) and magnetic field
  (0  - 33 T, $H$ $\parallel$ $c$) \cite{Zhu2002}.
  Optical conductivity $\sigma_1(E)$ was calculated by a Kramers-Kronig
  analysis \cite{Wooten1972}. Variable temperature
  transmittance was done on a 25 $\mu$m  $ab$ plane
  crystal, allowing direct calculation of absorption $\alpha(E)$.
    The Fe-57 M\"{o}ssbauer spectra of 35 mg/cm$^2$ of crushed crystals
    were recorded between 260 and 400 K on a constant
    acceleration spectrometer with a Rh matrix Co-57 source and
    calibrated at 295 K with $\alpha$-Fe powder.
    The reported isomer shifts are relative to $\alpha$-Fe at 295 K.
    X-ray scattering was performed at undulator beamline 4-ID-D at
  the Advanced Photon Source with  36 keV photons employing a
  cryostat with a split coil vertical field magnet up to 4 T.
  The sample was mounted with an angle of 45$^{\circ}$ between  $c$
 and the field as a compromise between cryomagnet  angular
restrictions  and accessibility of important regions of reciprocal
space. Previous magnetization work shows that
  LuFe$_2$O$_4$ is rather insensitive to
  fields  $\parallel$ $ab$ below 7 T. Consequently the
  dominate effect of the
 field is due to the component $\parallel$ $c$.


\begin{figure}[tb]
\centerline{
\includegraphics[width = 0.92\linewidth]{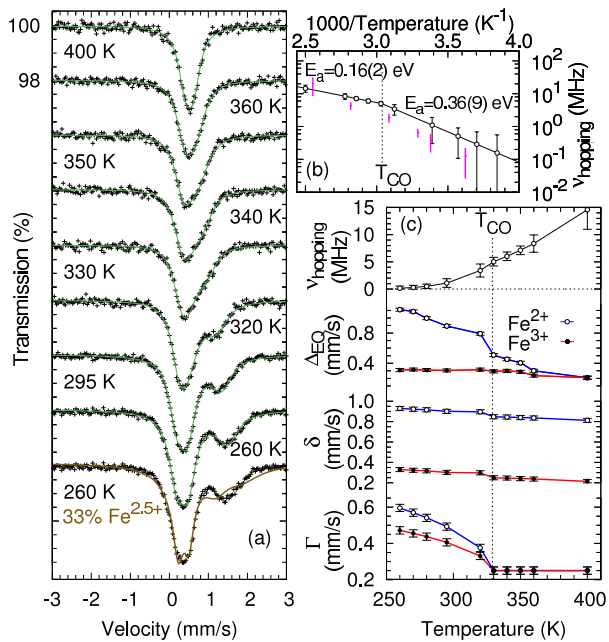}}
\caption{ \label{fig2_Mossbauer} (Color online) (a) M\"{o}ssbauer
spectra of LuFe$_2$O$_4$ and fits, see text; bottom: alternative
best fit with a constraint of 33\% of hopping electrons (b)
Arrhenius plot of the hopping frequency. Open symbols correspond to
the spectra in (a), bars to hopping frequencis in \cite{Tanaka1984}.
(c) M\"{o}ssbauer spectral parameters, from top to bottom: hopping
frequency, quadrupole splitting, isomer shift, and full linewidth at
half maximum. }
\end{figure}

The M\"{o}ssbauer spectra, (Fig.\ \ref{fig2_Mossbauer}(a)) were fit
with a Blume-Tjon model \cite{Tjon1968} for Fe$^{2+}$ and Fe$^{3+}$
relaxation, similar to \cite{Tanaka1984}. Individual fits of the
spectra reveal (i) two Arrhenius processes as indicated by the
temperature dependence of the hopping frequency and break at
$T_{CO}$ (Fig. \ref{fig2_Mossbauer}(b)), (ii) the difference between
Fe$^{2+}$ and Fe$^{3+}$ isomer shifts, $\Delta\delta$, is constant
below $T_{CO}$ and cannot be resolved above,
 and (iii) a constant linewidth and a gradual $\sim\sqrt{T_{CO}-T}$
 broadening above and below $T_{CO}$, respectively.
 In order to overcome correlation effects (see below) and to reduce the large number
 of fit parameters,
 a simultaneous parametric fit of all spectra was then carried
out with the constraints (i), (ii) constant $\Delta\delta$ for all
$T$, and (iii).
 The fits in Fig. \ref{fig2_Mossbauer}(a) are the result of this simultaneous fit.
The obtained spectral parameters and relaxation
frequencies (Fig. \ref{fig2_Mossbauer}(b,c)) are in agreement with 
\cite{Tanaka1984}, with the exception of the sharper $T$ dependence
of these parameters around $T_{CO}$, which we attribute to the
preparation of the sample as single crystal. Above $T_{CO}$, we
obtained an activation energy of 0.16(2) eV. This is somewhat
smaller than the 0.26 eV energy obtained by electrical conductivity
measurements \cite{Tanaka1984}. Below $T_{CO}$, we find the
activation energy to be 0.36(9) eV. The large error bar is due to
(i) small hopping frequencies, close to the detection limit, and
(ii) correlations between the hopping frequency and linewidth. The
Fe$^{2+}$ line around 1.4 mm/s is increasingly broadened below
$T_{CO}$ due to microscopic lattice distortions in the charge
ordered state  (Fig.\ \ref{fig2_Mossbauer}(a)) \cite{Tanaka1984}.
The fit at the bottom of Fig.\ \ref{fig2_Mossbauer}(a) is the best
fit with a constraint of 33\% of Fe$^{2.5+}$ with a hopping
frequency of 1 GHz. The poor fit quality  indicates that the 260 K
hopping frequency is smaller than 1 GHz, which invalidates the
presence of 33\% of Fe$^{2.5+}$ below $T_{CO}$ suggested in Ref.
\cite{zhang2007} from modeling electron diffraction, a technique
with a resolution  better than 1 GHz.

\begin{figure}[tb]
\centerline{
\includegraphics[width = 0.92\linewidth]{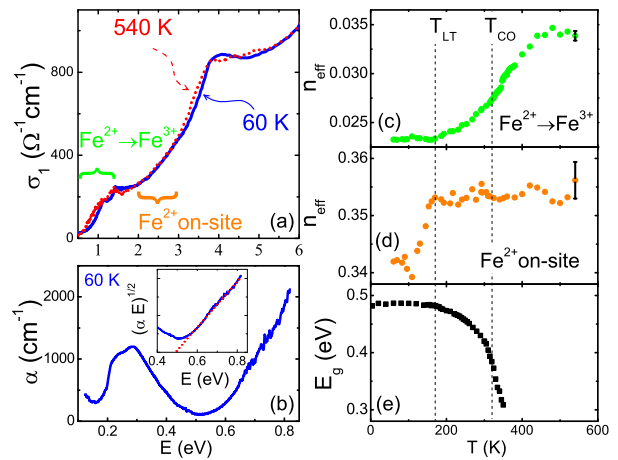}}
\caption{ \label{fig:optspectra} (Color online) 
Optical properties of LuFe$_2$O$_4$. (a) $\sigma_1$ vs energy $E$ at
540 and 60 K calculated from a Kramers-Kronig analysis of
reflectance. Brackets indicate assignments, the hierarchy determined
from Ref. \cite{Xiang2007}.
 (b) $\alpha$ vs $E$
calculated from transmittance, showing the charge transfer edge
starting from $\sim$0.5 eV. Inset: $\sqrt{\alpha E}$ with linear fit
(dotted line). (c,d) $n_{eff} (T)$ calculated by sum rules for
Fe$^{2+}\!\rightarrow$Fe$^{3+}$ charge transfer (c) and Fe$^{2+}$
on-site (d) excitations. (e) $E_g (T)$ calculated from an absorption
edge fit assuming an indirect gap. }
\end{figure}

To further study the charge fluctuation
(Fe$^{2+}\!\rightarrow$Fe$^{3+}$ charge transfer) observed in
M\"{o}ssbauer spectroscopy, we employed optical spectroscopy, a
technique in which hopping is driven by comparatively high frequency
photons ($\sim\! 10^{14}\, {\rm Hz}$, in contrast to the natural
hopping rate of a few MHz, as observed by M\"ossbauer spectroscopy).
Figure \ref{fig:optspectra}(a) displays the optical conductivity of
LuFe$_2$O$_4$. We assign the observed excitations based upon recent
first principle calculations \cite{Xiang2007}.
The lowest allowed electronic features  are minority channel
Fe$^{2+}\!\rightarrow$Fe$^{3+}$ charge transfer excitations. At
slightly higher energy follow Fe$^{2+}$ on-site excitations.
Minority channel O $p\!\rightarrow$Fe $d$ charge transfer and
overlapping majority channel O $p$ $\rightarrow$ Lu $s$ state
excitations are observed at higher energy ($\ge$3 eV). It is
difficult to resolve all the excitations because they are broad and
overlap significantly. Nevertheless,  the 1.1 eV peak in the near
infrared range (Fig.\ \ref{fig:optspectra}(a)) can be associated
with Fe$^{2+}\!\rightarrow$Fe$^{3+}$ charge transfer.
The optical gap $E_g$ is determined by the absorption edge of this
band. A close-up view of
the tail is shown in  Fig. \ref{fig:optspectra}(b). 
The linear fit of $\sqrt{\alpha E}$ vs. $E$ above 0.5 eV indicates
that the gap is indirect \cite{Wooten1972}, in agreement with Ref.
\onlinecite{XiangUnpub}.
Due to the 0.3 eV feature  (possibly a spin-forbidden Fe$^{3+}$
on-site excitation \cite{Henderson1989}), the typical ``double slope
character'' is not observed and the associated coupling phonon
energy for the indirect gap excitation process can not be
determined. The optical gap is sensitive to $T_{CO}$ (Fig.\
\ref{fig:optspectra}(e)), although LuFe$_2$O$_4$ is a semiconductor
(non-metallic) over the full temperature range of our investigation.

To quantify the strength of the various excitations, we calculated
the effective number of electrons $n_{eff}$ from the optical
conductivity $\sigma_1(\omega)$ using the partial sum rule:
$
n_{eff}\equiv\left.{\displaystyle\int_{\omega_1}^{\omega_2}\sigma_1(\omega)/\epsilon_0\,{\rm
d}\omega}\right/\displaystyle\frac{1}{2}\pi\omega_p^2 \ , $ where
$\omega_p \equiv \sqrt{\frac{e^2}{V_0m\epsilon_0}}$ is the plasma
frequency, $e$ and $m$ are the charge and mass of an electron,
$\epsilon_0$ is the vacuum dielectric constant, $V_0$ is the unit
cell volume, and $\omega_1$ and $\omega_2$ are the frequency limits
of integration. The absolute level of $n_{eff}$ depends on the
integration range, but the temperature trends  are not sensitive to
this choice.
For instance,  to investigate changes in the Fe charge transfer
band, we evaluated the partial sum rule from 0.6 - 1 eV, as
indicated in Fig.\ \ref{fig:optspectra}(a). In this case, $n_{eff}$
represents effective number of electrons that are able to overcome
the energy barrier to hop from Fe$^{2+}$ to Fe$^{3+}$ sites. This
number increases over a broad temperature range through $T_{CO}$, as
shown in Fig.\ \ref{fig:optspectra}(c).

In optical processes, electrons hopping from Fe$^{2+}$ to Fe$^{3+}$
are better described as small polarons, which correspond to combined
electronic and vibrational excitations  that arise
when the lattice is too slow to relax \cite{Yamada2000}. 
Important signatures include   (i) a large effective mass and (ii)
optical excitation
energies that are much larger than the low frequency activation energy. 
 The effective mass of the charge carriers can be
estimated using $n_{eff} = \frac{m}{m^*}N$, where $m^*$ is the
effective mass and $N=3$ (the number of Fe$^{2+}$ site per unit
cell). Considering that the 0.6 - 1 eV integration is only over half
of the excitation, we get $\frac{m^*}{m} \approx 40$, which is
large, but typical for polarons (e.g.\ $\frac{m^*}{m} \approx
100$ in Fe$_3$O$_4$ \cite{Gasparov2000}). With the polaron picture
and the simple model of an electron jumping between two sites
\cite{Austin1969}, we can estimate the 60 K activation energy from
the optical activation $\hbar\omega_0$ as
$E_a=\hbar\omega_0/4$=1.1(1)eV/4=0.28(3)eV,
 in excellent agreement with value reported in the study
of low frequency dielectric dispersion and DC electric conductivity
ranging from 0.25 - 0.29 eV \cite{Ikeda2005,Tanaka1984} and
compatible with that for spontaneous electron hopping obtained
between 260 and 320 K from M\"{o}ssbauer spectroscopy.

The $T$ dependence of $n_{eff}$ corresponding to Fe$^{2+}$ to
Fe$^{3+}$ charge transfer  confirms charge fluctuation below
$T_{CO}$ (Fig. 2(c)). Here, $n_{eff}$ begins to increase well below
$T_{CO}$ (evident also in  electron hopping trends via M\"{o}ssbauer
spectroscopy) and continues to change above this temperature. This
result is consistent with the
presence of 3D anti-phase domain boundary modes 
 \cite{Yamada1997}. We attribute the experimental observation of
significant charge fluctuations through $T_{CO}$ (even where
diffraction shows that it is ordered)
 to relevance of the ``order by fluctuation''
 mechanism in which fluctuations are needed to stabilize the
$\sqrt{3}$$\times$$\sqrt{3}$ CO in the frustrated system
\cite{Nagano2007}. Interestingly,  the charge fluctuation onset is
at  $T_{LT}$ (Fig. 2(c), (e)), suggesting that the low temperature
phase transition quenches the charge fluctuation. Similar to the
Verwey transition in Fe$_3$O$_4$ \cite{Park1998}, the $T$ dependence
of $n_{eff}$ shows an anomaly near $T_{CO}$ consistent with the
lowering of the activation energy above $T_{CO}$
\cite{Tanaka1984}. However, the jump occurs 
above $T_{CO}$.

\begin{figure}[tb]
\centerline{
\includegraphics[width= \linewidth]{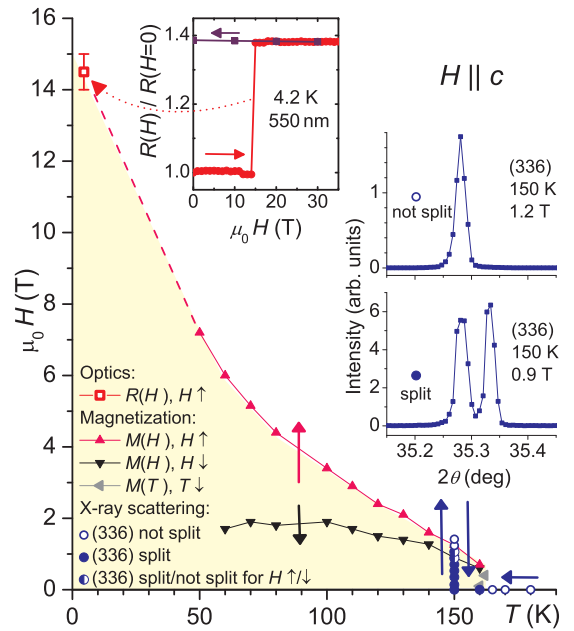}}
\caption{ \label{fig3} (Color online) Low-temperature phase
transition from magnetization data
($\blacktriangle\blacktriangledown\blacktriangleleft$,see
\cite{Christianson08}), field-dependence of optical reflectance
contrast $R(H)/R(H=0)$ at $\lambda$ = 550 nm (squares, see left
inset), and presence (indicated by $\bullet$) or absence ($\circ$)
of a splitting in the total diffraction angle of the structural
Bragg reflection (336) from synchrotron x-ray scattering (right
insets). The maximum region of the structurally distorted phase is
shaded (see text).}
\end{figure}

The strength of the Fe$^{2+}$ on-site crystal field excitation is
quantified by the partial sum rule in energy range 2 - 3 eV, in
accord with first principle calculations \cite{Xiang2007}. Although
the data are more scattered than that discussed above because of the
background from nearby excitations, two finding are immediately
clear. First, the Fe$^{2+}$ on-site excitation is rather insensitive
to charge and spin ordering transitions. Second, it displays a clear
anomaly  at $T_{LT}$ (Fig.\ \ref{fig:optspectra}(c)), which recent
magnetization and neutron diffraction studies \cite{Christianson08}
identified as an additional first-order transition. Further
magnetization measurements in $H\|c$ up to $7\,{\rm T}$ indicate
that the transition temperature is strongly suppressed by $H$ (Fig.\
\ref{fig3}). Hysteresis of $H_{LT} (T)$ widens upon cooling, and
below 50 K the high-$H$, high-$T$ phase remains frozen in even after
decreasing $H$ to $0$, reminiscent of kinetic arrest of first-order
transitions as studied, e.g., in doped CeFe$_2$ \cite{CeFe2}.

To check for possible \cite{Christianson08} structural components of
the transition, we closely examined structural Bragg reflections,
such as (336). CO below 320 K  \cite{Angst08} lowers the crystal
symmetry to monoclinic, which could lead to a monoclinic distortion
($\beta$ $\ne$ 90$^{\circ}$), with a splitting in $2\theta$ values
of such reflections, with domain formation clearly observable also
in single crystals \cite{Angst07}. Such a splitting could not be
resolved in between 200 and 300 K (in any magnetic field),
indicating that any monoclinic distortion is small. Splitting
becomes evident upon cooling below $T_{LT}$ (lower right inset in
Fig.\ \ref{fig3}), consistent with a significant ($\beta \approx
90.5^{\circ}$) monoclinic distortion \cite{ManuelNote2}. Application
of a magnetic field removes the splitting (upper right inset) as
soon as the field component $\|c$ reaches $H_{LT}$ as determined by
magnetization, and subsequent reduction causes a reappearance of the
splitting at  a field value again consistent with magnetization.
These diffraction data thus suggest that $T_{LT}$ and $H_{LT} (T)$
corresponds to a strongly hysteretic magneto-structural transition.
We propose that the monoclinic distortion removes geometric
frustration rendering charge fluctuation unnecessary. This scenario
is in line with the observed fluctuation onset at $T_{LT}$ and  the
``order by fluctuation'' mechanism.

The anomaly at $T_{LT}$ in the Fe$^{2+}$ on-site  excitations can
also be explained within a structural distortion scenario. Consider
an Fe center coordinated by five O ligands in a trigonal bipyramidal
geometry ($D_{3d}$ symmetry). A crystal field splits the Fe 3$d$
levels into three groups \cite{Nagano2007}. The monoclinic
distortion in the low $T$ phase splits these levels further,
shifting the on-site excitation energies and causing the
discontinuity in $n_{eff}$ (Fig. \ref{fig:optspectra}(d)).

The upper-left inset in Fig.\ \ref{fig3} displays the 4.2 K
reflectance of LuFe$_2$O$_4$ at 550 nm as a function of magnetic
field ($H$ $\parallel$ $c$). At this energy, the spectral response
is probing field-induced changes in Fe$^{2+}$ on-site excitations.
Strikingly, the reflectance increases by $\sim$40\% for $H$ $\ge$ 14
T. This 
is consistent with a
straight-forward extrapolation of $H_{LT}$ from the magnetization
data (dashed line in Fig.\ \ref{fig3}). That the reflectance
maintains this high value even after subsequent complete removal of
$H$ is again consistent with the magnetization data (see above).
Thus, at low $T$ a rather large field is required to switch the
crystal structure. Since the 
structural distortion is most likely induced by the CO, which lowers
the crystal symmetry \cite{Angst08}, the strong $H$-dependence of
this transition is a further example of the strong coupling between
spin, structural, and charge degrees of freedom in LuFe$_2$O$_4$.


In summary, optical and M\"{o}ssbauer  spectroscopies demonstrate
that charge fluctuation in LuFe$_2$O$_4$ has  an onset at $T_{LT}$,
well below  $T_{CO}$, supporting  the ``order by fluctuation''
mechanism for the $\sqrt{3}$ $\times$ $\sqrt{3}$ CO superstructure.
Fe$^{2+}$ on-site crystal field excitations are sensitive to the
magneto-structural transition at $T_{LT}$, which can be driven by
both temperature and magnetic field (requiring 14 T at 4 K).
Combining spectral, magnetic, and structural data, we generate a
comprehensive $H$-T phase diagram. The large temperature range of
the observed dynamical effects is a consequence of the  strong
coupling between  charge, structure, and magnetism.


We thank the Division of Materials Sciences and Engineering and the
Scientific User Facilities Division, Office of Basic Energy
Sciences, U.S. Department of Energy for support of this work at UT,
ORNL, and the APS. Work at the NHMFL is supported by NSF,  DOE, and
the State of Florida. Research at Li\`ege is funded by the FNRS. We
thank M.T. Sougrati for assistance with the M\"ossbauer and M.-H.
Whangbo and H. Xiang for useful discussions.


\begin{references}
\bibitem{ManuelNote2} The basic CO configuration appears the same with and without monoclinic
distortion \cite{Angst08}, but the magnetic structure is radically
changed \cite{Christianson08}.
\end{references}
\end{document}